\documentclass[physrev,twocolumn,amsmath,amssymb]{revtex4-2}

\usepackage{graphicx}
\usepackage[breaklinks,pdfborder={0 0 0},colorlinks=true,citecolor=blue,urlcolor=blue,linkcolor=blue]{hyperref}

\DeclareMathOperator{\Tr}{Tr}
\newcommand{\ket}[1]{|#1\rangle}
\newcommand{\bra}[1]{\langle#1|}
\newcommand{\mean}[1]{\langle#1\rangle}
\newcommand{\bmean}[1]{\bigl\langle#1\bigr\rangle}

\begin{document}

\title{Spatiotemporal Spread of Fermi-edge Singularity as Time Delayed Interaction and Impact on Time-dependent RKKY Type Coupling}

\author{Conor Jackson}
\author{Bernd Braunecker}%
\affiliation{%
SUPA, School of Physics and Astronomy, University of St Andrews, North Haugh, St Andrews KY16 9SS, United Kingdom
}%

\date{\today}

\begin{abstract}
Fermi-edge singularity and Anderson's orthogonality catastrophe are paradigmatic examples of
non-equilibrium many-body physics in conductors, appearing after a quench is created by the sudden change of a localised potential.
We investigate if the signal carried by the quench can be used to transmit a long ranged interaction,
reminiscent of the RKKY interaction, but with the inclusion of the full many-body propagation over space and time.
We calculate the response of a conductor to two quenches induced by localised states at different times and locations.
We show that building up and maintaining coherence between the localised states is possible only with finely tuned interaction between
the localised states and the conductor. This puts bounds to the use of time controlled RKKY type interactions and may
limit the speed at which some quantum gates could operate.
\end{abstract}

\maketitle

\section{Introduction}

\begin{figure}[t]
	\includegraphics[width=\columnwidth]{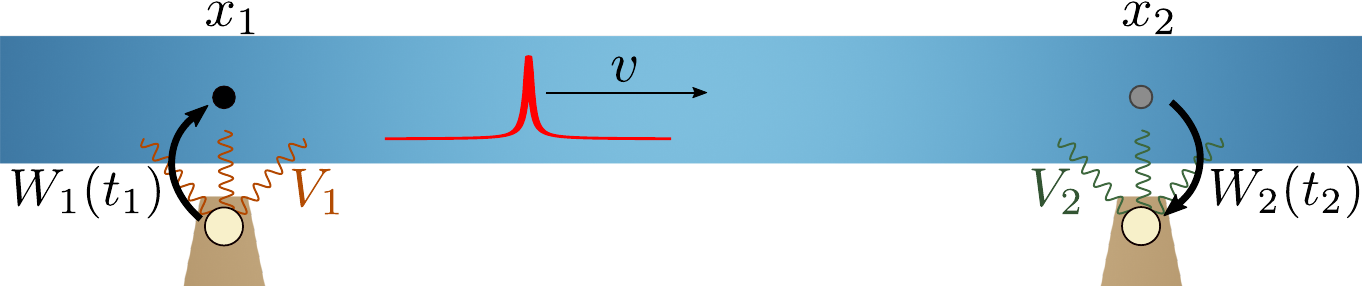}
	\caption{\label{fig:system}%
	Sketch illustrating the time non-local interaction mediated by a fermionic conductor,
	triggered by the tunnelling transitions
	$W_j(t_j)$ at positions $x_j$ and times $t_j$ for $j=1,2$. The same
	transitions switch on the FES scattering potentials $V_j$ renormalising the
	excitation peak travelling at velocity $v$.
	}
\end{figure}

The Fermi-edge Singularity (FES) Problem \cite{Mahan1967a,Roulet1969,Nozieres1969,Nozieres1969a} and Anderson's
Orthogonality Catastrophe (OC) \cite{Anderson1967} are the concepts behind one of the first and most important
examples of how a quench can drive a strongly correlated quantum response of a fermionic conductor.
For the OC the quench is caused by abruptly switching a localised scattering potential producing a proliferation
of zero energy particle-hole excitations. For the FES this is accompanied by the injection of an extra fermion into
the conduction band, or its extraction.
In both cases a screening cloud builds up near the potential that in the long time limit settles to a new ground state that is,
up to the extra fermion for the FES, orthogonal to the initial ground state. The relaxation of the overlap of initial and
final ground states follows a characteristic power-law in time that depends only on the potential's scattering phase shift.
Such a time dependence resembles the universal power-law responses of strongly correlated systems and makes FES/OC a model system for
quantum critical behaviour in the time domain. The Kondo effect in particular can be viewed as
a superposition of OC cascades triggered by the Kondo spin flips \cite{Yuval1970}.
Its universal many-body behaviour has made FES an important testing ground for a multitude of many-body techniques
over more than 50 years
\cite{Anderson1967,Mahan1967a,Nozieres1969,Roulet1969,Nozieres1969a,Schotte1969,Yuval1970,
Combescot1971,Ohtaka1990,Gogolin1993,Komnik1997,Muzykantskii2003,Abanin2004,DAmbrumenil2005,Braunecker2006,Bettelheim2006,Tureci2011,Snyman2013,Snyman2017}.
Experimental and theoretical evidence started with absorption and emission spectra in metals and semiconductors
\cite{Haensel1969,Tilton1974,Callcott1977a,Ishii1977,Flynn1985,Wertheim1992,Jiang1994,Adamjan1995,Swarts2005},
and extended then to nanostructured systems
\cite{Schmitt-Rink1986,Calleja1991,Matveev1992,Geim1994,Cobden1995,Itskevich1996,Benedict1998,Hapke-Wurst2000,Hentschel2005,Liu2006,Wang2006,Ruth2008,Latta2011,Heyl2012,Ubbelohde2012,
Chernii2014,Krahenmann2017,Goremykina2017,Ponomarenko2017,Kuhne2019,Ponomarenko2019}
and atomic gases
\cite{Knap2012,Sindona2013,Dora2013,Campbell2014,Schiro2014,Liu2019}.

It is, however, notable that with few exceptions \cite{Sheikhan2012,Snyman2014} the focus has been on global response functions,
and that there is a major lack of investigation of the spatial build-up and spread of the FES.
In this paper we show that the spatio-temporal spread offers a new perspective on FES physics,
and we provide a systematic access.
The FES quench can indeed be viewed as a coherent signal propagating through a fermionic bath. Picking up the signal
at some distance causes a coupling with the source of the quench. We formalise this aspect and
formulate the FES signal as a time delayed, long ranged effective interaction with a strong memory effect due to the
slow power-law decay of response functions. The memory effect invalidates the use of an effective Hamiltonian
so that we provide two appropriate formulations. One is fundamental in the form of a time dependent
action on the Keldysh contour that incorporates the concept of time delayed interaction.
The other one is formulated in the language of open system dynamics and
concretely focuses on the time evolution of the density matrix. The focus on the density matrix is motivated by the requirements of
quantum information processing.
We illustrate the approaches
through the example sketched in Fig.\ \ref{fig:system} where we investigate FES in combination with a quantum gate
operation between two localised qubit type states.
For concrete realisations one could consider a quantum wire coupled to two quantum dots or a fermionic atomic gas
extending across two trapped charges.
To transmit the signal we choose to inject a fermion into the conductor and extract it at a different location as this
represents the simplest case of such a transmission that captures the relevant physics.

The time delay in the signal is due to the finite Fermi velocity $v_F$ and does not require the FES itself. The latter, however,
causes a significant renormalisation of the transmission amplitude and decoherence even after
extraction unless special fine tuned conditions are met. This is in contrast the usual modelling of effective
interactions carried through a different medium such as the RKKY interaction. These are permanently present
and thus time and FES are of no significance. But if they are intended to be switched on and
off as required for a quantum gate operation \cite{Loss1998} our paper shows that time delay,
entanglement with the conductor and
the FES are essential processes to be taken into account.

The structure of the remaining paper is as follows.
In Sec.\ \ref{sec:model} we introduce the model represented in Fig.\ \ref{fig:system}.
In Sec.\ \ref{sec:action} we derive the time dependent action providing the conceptual picture for
the time delayed interaction. Section \ref{sec:rho} contains the concrete calculation of the
spatio-temporal response and its discussion. We conclude in Sec.\ \ref{sec:conclusions}.
Our results follow from calculations of a considerable length, as a side effect
of such a mature research topic. To keep this work accessible we concentrate in the main text
on presenting the physics of the spatio-temporal characteristics of FES with the discussion of its
consequences, and we leave the proper calculations in the background. We thus use the appendices to
provide the necessary methodologic and calculational details underpinning the discussion in a self-contained way.
Appendix \ref{sec:path_integral} contains the derivation of the path integral.
In Appendix \ref{sec:U(t)} we derive the shape of the evolution operator under the applied
pulses, and in Appendix \ref{sec:pulses} we evaluate it explicitly.
The structure of the amplitudes in the density matrix is derived in Appendix \ref{sec:structure_rho},
and in Appendix \ref{sec:bosonisation} we compute these amplitudes through the bosonisation method.


\section{Model}
\label{sec:model}

The minimal model shown in Fig.\ \ref{fig:system}
consists of a Fermi gas with two localised states.
For simplicity we consider spinless or spin polarised fermions, and we start with a noninteracting Hamiltonian
\begin{equation}
	H_c = \sum_k \epsilon_k c_k^\dagger c_k,
\end{equation}
where $\epsilon_k$ is the dispersion and $c_k$
are the fermion operators, but an extension to interactions will be considered later.
The localised states are single orbitals located at positions $x_j$
with operators $d_j$, energies $E_j$, and Hamiltonians
\begin{equation}
	H_j = E_j d_j^\dagger d_j,
\end{equation}
for $j=1,2$.
Transitions are induced by tunnelling terms
\begin{equation}
	H_W = \sum_j W_j(t) d_j^\dagger \psi(x_j) + \text{h.c.},
\end{equation}
where $\psi(x)$ is the field operator corresponding to $c_k$, and $W_j(t)$ are time dependent
amplitudes, applied over a range $\delta x_W \ll \pi/k_F$ (with Fermi momentum $k_F$)
such that tunnelling can be expressed as point-like at $x_j$.
The FES physics arises from the interaction
\begin{equation}
	H_{V} = V_1 \psi^\dagger(x_1) \psi(x_1) d_1 d_1^\dagger + V_2 \psi^\dagger(x_2) \psi(x_2) d_2^\dagger d_2.
\end{equation}
Here we assume that initially level $d_1$ is occupied and $d_2$ empty such that tunnelling out of $d_1$ and
into $d_2$ switches on the scattering potentials $V_j$. Tunnelling events are induced by
sharp pump peaks $W_j(t) = W_j \delta(t-t_j)$ that trigger the FES and make the concept of
a time delayed interaction between times $t_1$ and $t_2$ well defined. Such an operation comes also closest to
controlling an interaction between the $d_j$ levels as required for a quantum gate operation.
In the present case this could take the form of a conductor coupled to two quantum dot or defect states
that are pulsed to induce injection and readout.
We thus treat $H_W$ separately from the evolution
under $H = H_c+H_1+H_2+H_{V}$.


\section{Time non-local interactions}
\label{sec:action}

To illustrate how FES appears in this interaction we consider the effective action between the $d_j$ levels obtained in
the standard way from integrating out the conductor's
degrees of freedom in a path integral formulation.
To focus entirely on the physical interpretation we provide this derivation in Appendix \ref{sec:path_integral}.
For noninteracting fermions
the action is quadratic in the $\psi(x)$ fields and the path integral is readily evaluated.
This leads, in addition to the bare propagation under
the $H_{j}$, to the effective action $S_\text{eff} = S_{L} + S_{C}$ with (setting $\hbar=1$ throughout)
\begin{equation} \label{eq:S_L}
	S_L = \sum_{j,j'}\int_\mathcal{K} dt \, dt' W_j(t) d^\dagger_j(t) G(x_j,t;x_{j'},t') W_{j'}(t') d_{j'}(t'),
\end{equation}
and
\begin{equation}
	S_C = -i \Tr\ln(G_c G^{-1}),
\end{equation}
which are derived in Eqs.\ \eqref{eq:SC_app} and \eqref{eq:SL_app}.
Here $G$ is the full fermion propagator in the presence of a path
of the $d_j$ fields and $G_c$ the propagator for $d_j \equiv 0$.
In the FES language of Ref.\ \cite{Nozieres1969a}
$S_L$ is the open line propagator and $S_C$ is the
closed loop sum expressing the OC, which both are further resolved in space and include the two scattering centres $j=1,2$.
The full time dependence is retained, and the time integrations run
over the Keldysh contour $\mathcal{K}: -\infty \to +\infty \to -\infty$.
The trace in $S_C$ involves integration over contour time and space.

The $W_j(t)$ control the paths $d_j(t)$ in $S_\text{eff}$ and thus the response to this interaction.
Different paths of the $d_j(t)$ on the Keldysh branches $\mathcal{K}_\pm: \mp \infty \to \pm \infty$
encode all possible FES scenarios. Notable first is the absence of OC for classical realisations of $d_j(t)$,
which are equal on $\mathcal{K}_\pm$, and thus forward and backward time integrations are identical,
akin to the linked cluster theorem. Any interesting FES/OC effect is thus quantum with
different $d_j(t)$ on $\mathcal{K}_\pm$. The overlap integrals of FES spectra or of Loschmidt echos are
extreme examples with $V_j(t)$ nonzero only on one branch.
More general amplitudes involve the superposition of different paths, and
a richer example of quantum interference, involving newly the spatio-temporal response too, is the off-diagonal
density matrix element calculated below.

We observe that the time delayed interaction is carried only through the Green's function $G(x_j,t;x_{j'},t')$
in $S_L$, the open line contribution, as a consequence of the tunnel type interaction. This would be replaced
by two-fermion propagators for density-density interactions in $H_W$ instead, but does not involve the OC which
always acts as a ground state shake-up. In a fermionic liquid the $(x,t)$ dependence of Green's functions is
dominated by the Fermi edge cutoff which causes the characteristic power-law decay. The peak of these power-laws travels
with $v_F$ and thus causes largest impact at a distance $x$ only at time $x/v_F$. The interaction
$V_j$ modifies the power-law exponents through the characteristic phase shifts $\delta_j$.
Thus the interaction described by $S_\text{eff}$ combines three effects: the finite velocity of the interaction peak
which does not involve FES physics, the renormalisation of the peak shape by FES which can enhance or weaken the
signal, and the general decay by the OC which is always detrimental unless its long time behaviour can be switched
off by fine tuning. To substantiate these observations we consider now a concrete calculation.


\section{Time evolution of density matrix}
\label{sec:rho}

We illustrate the impact of spatio-temporal FES on the reduced density matrix $\rho_d$ of the two $d_j$ states
as a quantity relevant for quantum information processing that is accessible e.g.\ through quantum state
tomography.
Due to $W_j \delta(t-t_j)$ all times $t,t'$ in the path integral are pinned to $t_j$ on the Keldysh branches $\mathcal{K}_\pm$.
The number of possible paths $d_j(t)$ is then small and the path integral is evaluated directly.
It is advantageous to delay tracing out the $\psi(x)$ and write
$\rho_d(t) = \Tr_c\{ U(t) \rho(0) U^\dagger(t)\}$, where $\rho(0)$ is the full initial density matrix,
$U(t)$ the evolution operator and $\Tr_c$ the trace over the conductor's degrees of freedom.
In this way we rewrite the conceptual path integral in the language of open system dynamics of a reduced
density matrix $\rho_d$. For the pulsed transitions the number of paths is small and in this formulation we can evaluate
the contribution from each path directly.
The placement of $U$ and $U^\dagger$ corresponds to the evolution on the two
Keldysh branches.
For the pulsed transitions
the evolution operator takes the form \cite{Calkin1987,Nedeljkov2012}
\begin{equation} \label{eq:U}
	U(t) = e^{-i H_0 t} e^{-i \hat{W}_2} e^{-i \hat{W}_1}
\end{equation}
at $t>t_2>t_1$, where
\begin{equation}
	\hat{W}_j = e^{-i H_0 t_j} [ W_j d_j^\dagger \psi(x_j) + \text{h.c.}] e^{i H_0 t_j}.
\end{equation}
The effect of the pulses is thus entirely contained in the unitary operators $e^{-i \hat{W}_j}$,
and in Appendix \ref{sec:U(t)} we provide a discussion of the derivation of Eq.\ \eqref{eq:U}.
It is easy to show furthermore (see Appendix \ref{sec:pulses}) that
\begin{equation}
	e^{-i \hat{W}_j} = \openone - i \alpha_j \hat{W}_j - \beta_j \hat{W}_j^2,
\end{equation}
with $\alpha_j = \sin(w_j)/w_j, \beta_j = [1-\cos(w_j)]/w_j^2$
and $w_j \sim W_j/\sqrt{\delta x_W}$.
This means that the exact result takes the form of a second order perturbative expansion
but with renormalised coefficients $\alpha_j$ and $\beta_j$. This is a direct consequence of
fermion statistics and has the advantage that the number of paths $d_j(t)$ created remains
small such that the full time evolution can be evaluated exactly.

The two short pulses are chosen by analogy with the
switching on and off of an exchange interaction between two qubits as the basis of a quantum gate
generating entanglement between two qubits \cite{Loss1998}. For the transmitted signal we obtain a similar
gate operation between the $d_j$ levels but also decoherence due to the continuum of the fermionic fluctuations,
both being strongly affected by the FES. To illustrate we assume that initially $\rho_d(0) = \ket{1,0}\bra{1,0}$,
where $\ket{n_1,n_2}$ is the occupation number basis for $n_j=0,1$ the eigenvalues of $d_j^\dagger d_j$.
Applying a pulse on $d_1$ at $t_1$ followed by a pulse on $d_2$ at $t_2$ the density matrix at time $t>t_2>t_1$
takes the form
\begin{equation} \label{eq:rho_d}
	\rho_d(t)
	= \begin{pmatrix}
		D_0 & 0  & 0   & 0  \\
		0   & A  & C^* & 0  \\
		0   & C  & B   & 0  \\
		0   & 0  & 0   & D_1
	\end{pmatrix},
\end{equation}
spanned in the basis $\{ \ket{0,0}, \ket{1,0}, \ket{0,1}, \ket{1,1} \}$. The zeros arise from terms without particle
conservation along the full Keldysh contour.

Initially $A=1$ and all other amplitudes are zero. To put the result in context we compare it with
the effect of a conventional exchange interaction $H_J = J (d_1^\dagger d_2 + d_2^\dagger d_1)$.
Since $H_J^3 = J^2 H_J$ the evolution operator
becomes $U_J(t)= e^{-i H_J t} = \openone - i \sin(J t) H_J - [1-\cos(J t)] H_J^2 / J$, similarly to
the form of $U(t)$. The resulting density matrix takes the same form as Eq.\ \eqref{eq:rho_d}
with $A=\cos(Jt), B=\sin(Jt), C= i \sin(2Jt)/2$,
and $D_0=D_1=0$. The density matrix remains a pure state
but entangles the qubits.
Therefore, such an interaction can be used for a quantum gate, with the time of
interaction $t$ being the control parameter.

The present case for $\rho_d$ has several crucial differences due to the coupling to a
fluctuating fermionic continuum. Primarily it makes the evolution of $\rho_d(t)$ non-unitary, by
the creation of entanglement with the continuum that is lost for entangling the qubits.
This naturally causes decoherence but also the time retardation features by the FES and the
interaction that are our focus here.
These are illustrated through
the coefficient $C$, expressing the only route to build up entanglement between the two $d_j$ states.
The other coefficients involve only classical paths $d_j(t)$
as they are on the diagonal of $\rho_d(t)$ and thus must have identical sequences of $\hat{W}_j$
in the evolution operators to the left and right of $\rho_d(0)$, which means identical sequences on
both branches of the Keldysh contour. As noted above this cancels the OC response and as a consequence these
coefficients show no or only a weak FES dependence.
On the other hand, coefficient $C$ is off diagonal and requires different numbers of operators $\hat{W}_j$
on the two Keldysh branches. Namely we have
\begin{align}
	C &= - \alpha_1 \alpha_2 \,
		\bra{1,0} \Tr_c\bigl\{
			e^{-i H t}
			\bigl(\openone - \beta_2 \hat{W}_2^2 \bigr)
			\bigl(\openone - \beta_1 \hat{W}_1^2 \bigr)
\notag\\
	&\quad\ \times
			\rho(0)
			\hat{W}_1 \hat{W}_2
			e^{i H t}
		\bigr\} \ket{0,1}.
\label{eq:C}
\end{align}
The detailed derivation of this expression as well as expressions for the other coefficients
is provided in Appendix \ref{sec:structure_rho}.
Equation \eqref{eq:C} is exact but the contributions in terms of $\beta_{1,2}$ provide
only qualitatively similar corrections to the leading term so that in the discussion we shall focus
on the leading expression only, whereas in the figures we plot the full expression.
Since the operators on both Keldysh branches, to the left and right of $\rho(0)$ in Eq.\ \eqref{eq:C},
are different both FES and OC persist.
With the $\hat{W}_1\hat{W}_2$ operators on the right to the
initial density matrix $C$ takes indeed a form similar to the
Loschmidt echo, characterised by the interference of the perturbed system with the
free evolution, with the difference that at time $t_2$ instead of switching off the interaction
a tunnelling event takes place and the system evolves further to time $t$ (notice also that
$C=0$ for $t<t_2$).

\begin{figure}
	\centering
	\includegraphics[width=\columnwidth]{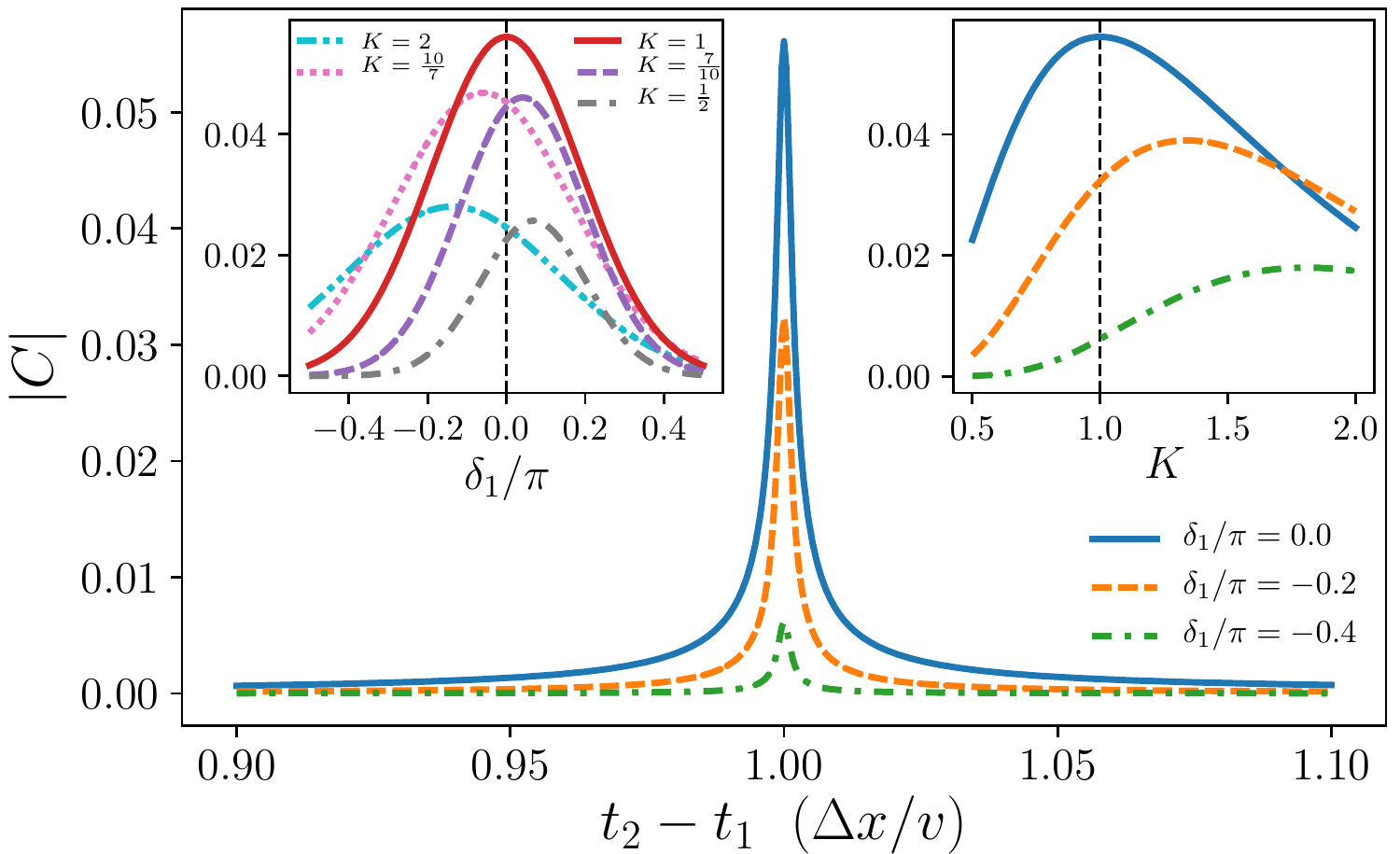}
	\caption{\label{fig:C_vs_t2}%
	Plot of coefficient $|C(t=t_2)|$ at $K=1$ for the indicated phase shifts $\delta_1$
	for repulsive $V_j<0$ at $\Delta x = x_2-x_1 = 1000 a$ (using $a=v_F=1$ to set space and time units).
	The finite time-of-flight causes the sharp peak
	at $t_2-t_1 = \Delta x/v$, and the FES its substantial suppression.
	The insets show the further suppression of the
	peak amplitude with interactions $K$ (right), and with $\delta_1$ for a selection of $K$ (left).
	Used parameters are $W_j=0.8 \sqrt{a}$, $n_p=0.4/a$, $n_h=0.6/a$, $\delta x_W=a$.
	}
\end{figure}%

Since three times $t_1,t_2,t$ and two positions $x_1, x_2$ are involved
the number of correlation functions is large for each amplitude in $\rho(t)$ but an analytic evaluation is
possible. To maximise the amplitude of the transmitted interaction we consider a one-dimensional (1D)
conductor in which excitations can run only to the left or the right. This allows us in addition to use the
bosonisation technique which provides the most straightforward technique to access the FES
physics \cite{Schotte1969,Gogolin1998,Giamarchi2007,Shankar2017}
and allows us to include interactions in the 1D conductor as well. In 1D the interactions are
entirely characterised through a parameter $K$ such that in the noninteracting case $K=1$, for
repulsive interactions $0<K<1$, and for attractive interactions $K>1$. Considering the zero temperature limit
the coefficients of $\rho$ are then composed of power laws expressing the propagation of the fermion from
$(x_1,t_1)$ to $(x_2,t_2)$. Through the FES the power-law exponents depend on the phase shifts
$\delta_j = 2 V_j K /v$ induced by the scattering potentials $V_j$ \cite{Schotte1969}.
Here $v=v_F/K$ is the interaction renormalised Fermi velocity \cite{Gogolin1998,Giamarchi2007,Shankar2017}.
The detailed evaluation is a standard calculation of some length but does not contribute further to the discussion.
We thus provide the details in Appendix \ref{sec:bosonisation} and analyse here instead the physics resulting from
the spatio-temporal spread.
We should notice though that we do not consider backscattering on the $W_j$ potentials.
Although this scattering is relevant in 1D it matters for weakly interacting systems
mostly at time scales that can be tuned to be longer than the times considered here, and its
inclusion would unnecessarily obscure the results.
Further technical but also more quantitative arguments are given in Appendix \ref{sec:bosonisation}.

To leading order in the power laws $C$ is expressed as
\begin{align}
	C
	&=
	-\alpha_1 \alpha_2 W_1 W_2
	e^{-i \Delta E_1 (t_1-t)}
	e^{-i \Delta E_2(t_2-t)}
\notag\\
	&\times
	\bigl( e^{-i k_F(x_1-x_2)} C_- + e^{+i k_F(x_1-x_2)} C_+ \bigr) / 2\pi a,
\end{align}
where $a$ is the short distance cutoff, and
where the coefficients $C_{\nu}$ with $\nu=\pm$ arise from the injection of a
right or left moving fermion with momenta near the Fermi momentum $\nu k_F$.
If we introduce $g_{x,t} = (a - i x + i v t)/a$ as the power-law basis
we have [see Eq.\ \eqref{eq:C_nu_full}]
\begin{align}
	&C_\nu
	=
	g_{0,       t_1-t}^{   \frac{\delta_1}{\pi K}-\frac{2\delta_1^2}{\pi^2 K}}
	\
	g_{0,       t_2-t}^{  -\frac{\delta_2}{\pi K}-\frac{2\delta_2^2}{\pi^2 K}}
	g_{x_1-x_2, 0}^{      -\frac{\nu \delta_1}{\pi} + \frac{\delta_1 \delta_2}{\pi^2 K}}
	\
	g_{x_2-x_1, 0}^{       \frac{\nu \delta_1}{\pi} + \frac{\delta_1 \delta_2}{\pi^2 K}}
\notag\\&\times
	g_{x_1-x_2, t_1-t_2}^{-\frac{(1+K \nu)^2}{4 K}+\frac{(\delta_1-\delta_2)(1+K \nu)}{2\pi K} + \frac{\delta_1 \delta_2}{\pi^2 K}}
	g_{x_1-x_2, t_1-t}^{   \frac{\delta_2(1+K \nu)}{2\pi K} - \frac{\delta_1 \delta_2}{\pi^2 K}}
\notag\\&\times
	g_{x_2-x_1, t_1-t_2}^{-\frac{(1-K \nu)^2}{4 K}+\frac{(\delta_1-\delta_2)(1-K \nu)}{2\pi K} + \frac{\delta_1 \delta_2}{\pi^2 K}}
	g_{x_2-x_1, t_1-t}^{   \frac{\delta_2(1-K \nu)}{2\pi K} - \frac{\delta_1 \delta_2}{\pi^2 K}}
\notag\\&\times
	g_{x_1-x_2, t_2-t}^{  -\frac{\delta_1(1-K \nu)}{2\pi K} - \frac{\delta_1 \delta_2}{\pi^2 K}}
	g_{x_2-x_1, t_2-t}^{-\frac{\delta_1(1+K \nu)}{2\pi K} - \frac{\delta_1 \delta_2}{\pi^2 K}}.
\label{eq:C_full}
\end{align}
The multitude of power laws in Eq.\ \eqref{eq:C_full} expresses the main result of this paper. It describes the various ways
in which FES and OC shake up the conductor and propagate between the times $t_1, t_2$, and $t$, as well as
between the positions $x_1$ and $x_2$, and how the interaction with $K \neq 1$ causes further fractionalisation
of the charge excitations. To understand the result let us consider first $t=t_2$, at which all dependence on
$\delta_2$ vanishes as $V_2$ acts only for $t>t_2$. We then have
\begin{align}
	&C_\nu
	=
	g_{0,       t_1-t_2}^{   \frac{\delta_1}{\pi K}-\frac{2\delta_1^2}{\pi^2 K}}
	g_{x_1-x_2, 0}^{  -\frac{\delta_1(1+K \nu)}{2\pi K}}
	g_{x_2-x_1, 0}^{-\frac{\delta_1(1-K \nu)}{2\pi K}}
\notag\\&\times
	g_{x_1-x_2, t_1-t_2}^{-\frac{(1+K \nu)^2}{4 K}+\frac{\delta_1(1+K \nu)}{2\pi K}}
	g_{x_2-x_1, t_1-t_2}^{-\frac{(1-K \nu)^2}{4 K}+\frac{\delta_1(1-K \nu)}{2\pi K}}.
\end{align}
For $x_2-x_1=0$ this expression reduces further to the standard FES response
whereas at nonzero $x_2-x_1$, as seen in Fig.\ \ref{fig:C_vs_t2}, the last two factors produce a pronounced peak at $t_2-t_1 = |x_2-x_1|/v$
as consequence of the finite propagation velocity.
Hence, in contrast to the instantaneous $H_J$ a fine tuning must be made to maximise the correlation between the two levels.
Note that since the Hamiltonian is nonrelativistic tails of the correlators build up immediately
and $C$ is nonzero already at all $t_2-t_1<|x_2-x_1|/v$.

Figure \ref{fig:C_vs_t2} also shows the substantial impact of FES.
Varying $\delta_1$ from $0$ to $\pm \pi/2$ suppresses the peak amplitude by
more than an order of magnitude,
but in the tails we observe that the interplay between the OC and the contribution from the added fermion
can lead to both larger or smaller amplitudes. The figure shows as well that interactions have a similar
reducing effect but display also a partial compensation of the FES effect by shifting the maximum
of $C$. Similar behaviour can be found in the dynamics of a Kondo spin coupled to interacting chains \cite{Braganca2021} and is thus not specific for 1D.

\begin{figure}
	\centering
	\includegraphics[width=\columnwidth]{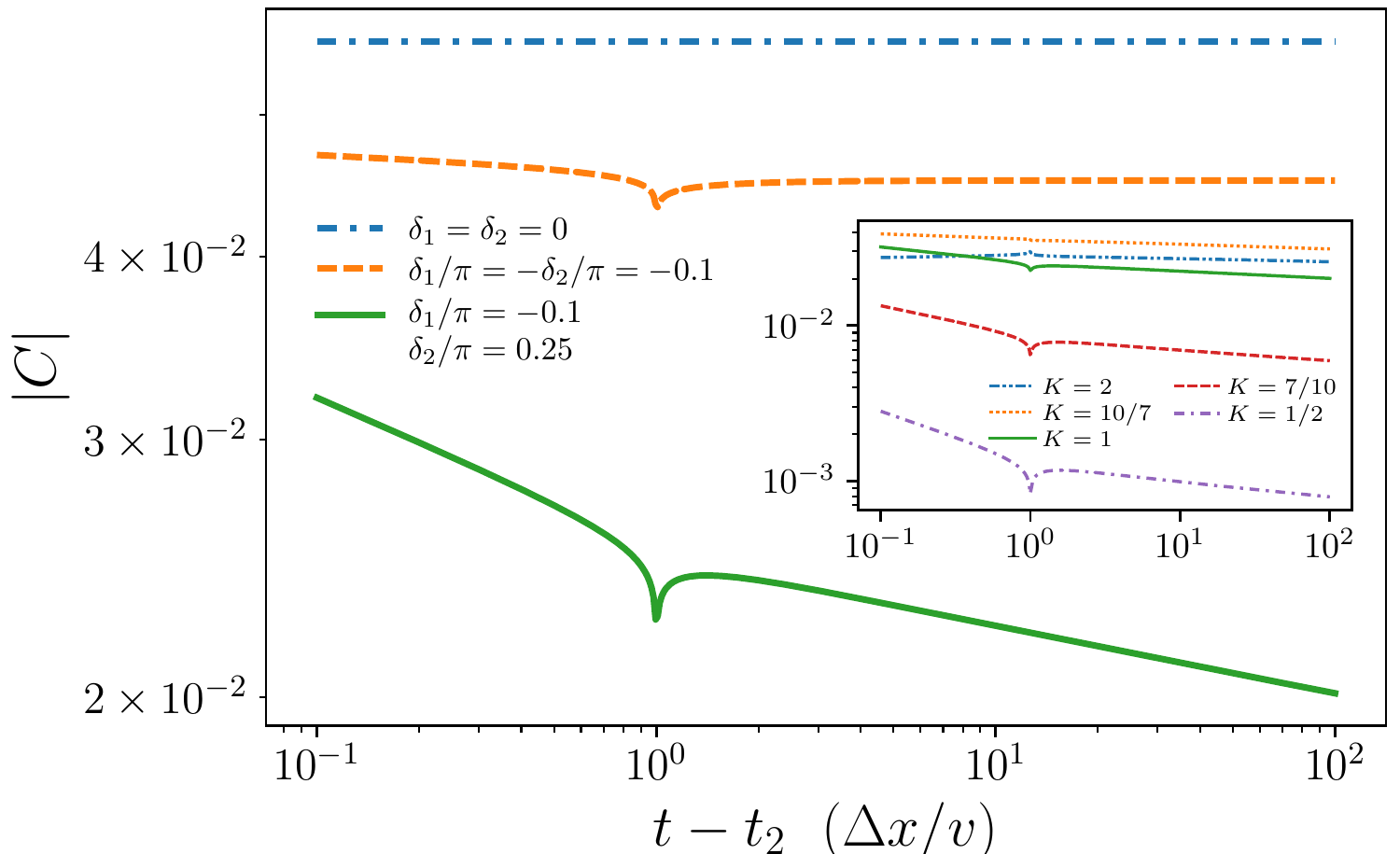}
	\caption{\label{fig:C_vs_t}%
	Plot of $|C(t)|$ at $t>t_2$ for $K=1$ and parameters as in Fig.\ \ref{fig:C_vs_t2}. For $\delta_1=-\delta_2$
	coefficient $|C|$ becomes constant at large $t$ but decays for all other $\delta_j$. The inset shows the dependence on
	interactions $K$ for $\delta_1/\pi=-0.1$ and $\delta_2/\pi=0.25$.
	}
\end{figure}%

Further distinction from a standard quantum gate
arises entirely from the OC induced relaxation.
This is the lasting effect of the correlation with the conductor and takes the role of
an interaction that cannot just be switched off unless further fine tuning is achieved.
Indeed $C \sim t^{-2(\delta_1+\delta_2)^2/\pi^2 K}$ at long times $t \gg t_2$, suppressing with $C \to 0$ thus the transfer of
information between the $d_j$  unless the $V_j$ are such that $\delta_1 = -\delta_2$.
The behaviour as function of $t>t_2$
is shown for a selection of phase shifts and $K=1$ in Fig.\ \ref{fig:C_vs_t} in which the
condition $\delta_1=\delta_2=0$ leaves $C$ unchanged from its magnitude at $t=t_2$, as is
indeed expected for a RKKY type coupling in a noninteracting system.
For $\delta_1=-\delta_2\neq 0$ the amplitude saturates as noted at a constant value at
large $t$. But the signal shows now a transient behaviour with a
satellite peak at $t-t_2=|x_2-x_1|/v$ caused by one of the last two factors in Eq.\ \eqref{eq:C_full} and relies on $\delta_1 \neq 0$.
This effect arises due to the many-body interference of the OC at $x_1$ with the excitations created by the fermion absorption at $(x_2,t_2)$, manifesting as another Fermi-edge-style singularity.
For $\delta_1+\delta_2 \neq 0$ we see a similar transient behaviour but then the further
(albeit rather slow) decay with increasing $t$. Interactions with $K \neq 1$ are shown in the inset
of Fig.\ \ref{fig:C_vs_t}. Here interactions can reduce and even enhance the
amplitudes but do not change the qualitative features.

The time and space resolved spread of FES described by Eq. \eqref{eq:C_full} therefore not just exhibits
what could be thought as an obvious extension of the well known FES results to a peak travelling at finite velocity $v$.
Instead all possible interference effects in time only, space only, and time and space mixed contribute
each with a characteristic power-law exponent.


\section{Conclusions}
\label{sec:conclusions}

The results above provide an extension to our understanding of FES/OC physics to how the excitation extends non-locally
through space and time. We illustrated the impact on response functions through the example of correlated tunnel events,
as a simple example of a quantum information type setup. With the finite propagation velocity the transition amplitudes
are strongly peaked at the characteristic run time, and the FES causes a reduction of the peak amplitude together with
the modification of the power-law tails. The OC generally remains detrimental to any long term quantum coherence, even
long after the second transition, unless the different scatterers are fine tuned to identical phase shifts. But even then
there is a transient further decay before matrix elements of $\rho_d$ converge to a nonzero constant, which is the general
result of the many interference processes in Eq.\ \eqref{eq:C_full} distinguishing this case from the standard global
FES/OC response.
As the transient regime contains the most interesting coherent correlations, experimental probing would require ultra-fast
techniques, e.g.\ for electron conductors with THz resolution. This could be offered by multidimensional
spectroscopy \cite{Kuehn2010,Maiuri2020,Lloyd-Hughes2021}, and could target the tomography of $\rho_d$ or the conductor's
excitations directly.
On the other hand if conductors are used for RKKY type interactions between quantum gates \cite{Trifunovic2012,Yang2016},
then sharp pulses triggering FES should be avoided and a smoother operation such as with minimal excitation
pulses \cite{Keeling2006,Dubois2013} should be chosen.
This would necessarily slow down the rate at which such a gate could be operated.
As a fundamental result, however, we have shown that the inclusion of a spatial component causes the FES response to split
into all possible interference combinations between the involved time and space variables. While the OC diagrammatically
remains decoupled from all other correlators and thus maintains its pure time dependence, all other FES processes connect
the variables in a rather nontrivial manner, yet all with characteristic power-laws.


\begin{acknowledgments}
We thank H. T\"{u}reci for a stimulating discussion.
C.J. acknowledges the support from the EPSRC under Grant No.\ EP/L015110/1.
The work presented in this paper is theoretical. No data were produced,
and supporting research data are not required.
\end{acknowledgments}

\appendix

\section{Path integral for interaction}
\label{sec:path_integral}

The derivation of the effective action for the time retarded interaction starts from the full action of the combined
system of localised $d_j$ states and the continuum $\psi(x)$. If these $d_j$ and $\psi(x)$ represent now
Grassmann fields the action reads $S = S_c + \sum_{j=1,2} S_j + S_V + S_W$, with
\begin{align}
	S_c = &\int_\mathcal{K} dt \int dx \, \psi^\dagger(x,t) \bigl(i \partial_t - H_c\bigr) \psi(x,t),
\\
	S_j = &\int_\mathcal{K} dt \, d_j^\dagger(t) \bigl(i \partial_t - E_j\bigr) d_j(t),
\\
	S_V = &- \int_\mathcal{K} dt \, V_1 \psi^\dagger(x_1,t) \psi(x_1,t) \bigl[1-d_1^\dagger(t) d_1(t)\bigr]
\notag\\&
	     - \int_\mathcal{K} dt \, V_2 \psi^\dagger(x_2,t) \psi(x_2,t) d_2^\dagger(t) d_2(t),
\\
	S_W = &\sum_{j=1,2} \int_\mathcal{K} dt \, W_j(t) \psi^\dagger(x_j,t) d_j(t) + \text{c.c.},
\end{align}
in which the time integrals run over the Keldysh contour $\mathcal{K}: -\infty \to +\infty \to -\infty$.
We consider only noninteracting fermions here in which $H_c$ is local and thus $S_c$ requires only a single spatial integral.

To obtain an effective action for the $d_j$ levels we integrate out the $\psi(x)$ fields as
\begin{align}
	e^{i S_\text{eff}[d_j^\dagger,d_j]} =
	\int D[\psi^\dagger,\psi] \, e^{i (S_c + S_V + S_W)},
\end{align}
such that the total effective action is $\sum_{j} S_j + S_\text{eff}$.
In the absence of interactions, as considered for the evaluation of the path integral, the $\psi(x)$ integrals are Gaussians
and the $\psi$ integration is straightforward,
\begin{align}
	\int D[\psi^\dagger,\psi] \, e^{i(\psi|G^{-1} \psi) + (b|\psi) + (\psi|b)}
	= \det(iG^{-1}) \, e^{i (b|G b)},
\end{align}
where the inner product $(\cdot|\cdot)$ consists of the $x$ and $t$ integrations.
We write the Green's function in the kernel as $G^{-1} = G_c^{-1} - V$, with
$G_c^{-1}(x,t;x',t') = \delta(t-t') \delta(x-x') [i \partial_t - H_c]$
and
$V(x,t;x',t') = \delta(t-t') \{V_1 \delta(x-x_1)\delta(x'-x_1) [1-d_1^\dagger(t)d_1(t)] + V_2 \delta(x-x_2)\delta(x'-x_2) d_2^\dagger(t) d_2(t)\}$,
and the inhomogeneous terms as
$b(x,t) = \sum_j W_j(t) \delta(x-x_j) d_j(t)$.

The determinant factor can be written in the form
\begin{align}
	\det(iG^{-1})
	&= \det(i G^{-1}_c) \, \exp\bigl(\Tr \ln(G_c G^{-1})\bigr).
\end{align}
Here $\det(i G^{-1}_c)$ is an unimportant constant that can be dropped.
On the other hand $e^{i S_C} = \exp(\Tr \ln(G_c G^{-1}))$ is important.
If we write it as
\begin{equation} \label{eq:SC_app}
	S_C = -i \Tr \ln(1 + i G_c V)
\end{equation}
we see from expanding the logarithm
that it describes the full
set of simple closed loop diagrams connecting vertices $V$. This term therefore
incorporates the closed loop contribution responsible for the OC \cite{Nozieres1969a}.
The full effective action then becomes $S_\text{eff} = S_L + S_C$ with
\begin{equation} \label{eq:SL_app}
	S_L = \sum_{j,j'} \int dt \, dt' \, W_j(t) d_j^\dagger(t)
	G(x_j, t; x_{j'}, t') W_{j'}(t') d_{j'}(t'),
\end{equation}
in which $G$ is the full Green's function on the Keldysh contour including the scattering on the
time dependent potential $V$ created by the realisations of the $d_j(t)$ fields. In contrast to
the closed loops in $S_C$ the propagator is pinned to the times $t,t'$ at which the pulses $W_j$
are active and thus $S_L$ generalises the open line diagrams of the FES \cite{Nozieres1969a}.


\section{Evolution operator for delta function pulses}
\label{sec:U(t)}

We consider a time dependent Hamiltonian of the form
\begin{equation} \label{sup:H}
	H_\text{full}(t) = H + W \delta(t-t_1),
\end{equation}
with $H = H_c + \sum_j H_j + H_V$ time independent and $W$ applied only through a pulse at time $t_1$
that is sharp enough to be treated as a delta function pulse. Such a time dependence allows
for a simple solution for the evolution operator $U(t)$ which, however, requires some care.
Indeed integrating the equation of motion
$i \partial_t U(t) = H_\text{full}(t) U(t)$ over times
$t_1-\delta t < t < t_1+\delta t$ for some $\delta t>0$
produces $U(t_1+\delta t)-U(t_1-\delta t) = -i W U(t)$. The fact that the right hand side is nonzero
shows on the left hand side that $U(t)$ is discontinuous at $t_1$. This in turn makes the right hand
side ambiguous. The correct treatment of this situation is an old problem and, for instance,
in Refs.\ \cite{Calkin1987,Nedeljkov2012} a thorough discussion is provided.

It turns out that the naive solution produces the correct answer. If we solve the equation of motion in the usual
way by going to the interaction picture with $W$ as perturbation and perform a formal integration we obtain
the standard form of the time ordered exponential
\begin{equation} \label{sup:U_formal}
	U(t) = e^{-i H t} T \exp\left(-i \int_0^t dt' \, \hat{W}(t') \delta(t'-t_1)\right),
\end{equation}
where $T$ is the time ordering operator and $\hat{W}(t) = e^{-i H t} W e^{i H t}$.
The implicit but far from obvious assumption in Eq.\ \eqref{sup:U_formal} is that $T$ commutes with the integration.
Accepting it though allows us immediately evaluate the integral in the exponential. Noting then that $T$ has no
effect for equal time expressions we obtain
\begin{equation} \label{sup:U_sol}
	U(t) = e^{-i H t} e^{-i \hat{W}(t_1)} = e^{-i H (t-t_1)} e^{-i W} e^{-i H t_1},
\end{equation}
for $t>t_1$ and $U(t)=e^{-iH t}$ for $t<t_1$.
Although a rigorous treatment requires a more refined approach \cite{Calkin1987,Nedeljkov2012},
Eq.\ \eqref{sup:U_sol} is indeed the correct result. The last part of the equation provides the appropriate
physical picture: the system evolves under $H$ before and after $t_1$, and the effect of the pulse is entirely
contained in the unitary and nonperturbative operator $e^{-iW}$.

From the latter expression it is straightforward to obtain the evolution operator for sequences of
pulses. Considering two pulses, $H_\text{full}=H + W_1 \delta(t-t_1)+ W_2 \delta(t-t_2)$, we have
\begin{align}
	U(t)
	&= e^{-i H (t-t_2)} e^{-i W_2} e^{-i H (t_2-t_1)} e^{-i W_1} e^{-i H t_1}
\notag\\
	&= e^{-i H t} e^{-i\hat{W}_2(t_2)} e^{-i \hat{W}_1(t_1)},
 \label{sup:U2_sol}
\end{align}
where we have assumed $t > t_2 > t_1$. Further pulses chain up in the same way.


\section{Evaluation of the pulse operators}
\label{sec:pulses}

The operators $e^{-i \hat{W}_j}$ in the evolution operator resulting from pulses at times $t_j$
can be given a closed form in which we only need to be careful with infinities.
The point-tunnelling expressions
\begin{equation} \label{sup:W_point}
	\hat{W}_j = e^{i H t_j} \bigl( W_j d_j^\dagger \psi(x_j) + \text{h.c.} \bigr) e^{-i H t_j}
\end{equation}
cause at higher powers in the expansion of $e^{-i \hat{W}_j}$ products of $\psi(x_j)$ and $\psi^\dagger(x_j)$
that through the anticommutation rule $\{\psi(x), \psi^\dagger(x')\} = \delta(x-x')$ cause divergences.
It thus must be noted that Eq.\ \eqref{sup:W_point} itself is only a convenient limit of the more general
interaction
\begin{equation} \label{sup:W_int}
	\hat{W}_j = e^{i H t_j} \int dx \bigl( W_j(x) d_j^\dagger \psi(x) + \text{h.c.} \bigr) e^{-i H t_j},
\end{equation}
where $W_j(x)$ is a spatially dependent potential that is sharply peaked at $x=x_j$ and integrates to the
amplitude $W_j$ in Eq.\ \eqref{sup:W_point}.
The order $\hat{W}_j^2$ itself would be unproblematic even with Eq.\ \eqref{sup:W_point}.
Using Eq.\ \eqref{sup:W_int} we find that
\begin{align}
	\hat{W}_j^2 &= e^{i H t_j} \int dx dx' \, W_j(x) W_j(x')
\notag\\
	&\quad\times
	\bigl(
		d_j^\dagger d_j \psi(x)\psi^\dagger(x')
		+
		d_j d_j^\dagger \psi^\dagger(x)\psi(x')
	\bigr)e^{-i H t_j}
\notag\\
	&\approx
	W_j^2
	e^{i H t_j}
	\bigl(
		d_j^\dagger d_j \psi(x_j)\psi^\dagger(x_j)
		+
		d_j d_j^\dagger \psi^\dagger(x_j)\psi(x_j)
	\bigr)
\notag\\
	&\quad\times
	e^{-i H t_j},
\end{align}
where the second line would also follow from Eq.\ \eqref{sup:W_point} and
can be used for the practical evaluation of $\hat{W}_j^2$.
For $\hat{W}_j^3$, however, we obtain
\begin{align}
	&\hat{W}_j^3
	= e^{i H t_j}\int dx dx' dx'' \, W_j(x) W_j(x') W_j(x'')
\notag\\
	&\times
	\bigl(
		d_j^\dagger d_j d_j^\dagger  \psi(x) \psi^\dagger(x') \psi(x'')
		+
		d_j d_j^\dagger d_j  \psi^\dagger(x) \psi(x') \psi^\dagger(x'')
	\bigr)
\notag\\
	&\times
	e^{-i H t_j}.
\label{sup:W3_1}
\end{align}
Through the anticommutation relations for the $\psi(x)$ and $d_j$ this
expression can then be reduced to
\begin{align}
	\hat{W}_j^3
	&= e^{i H t_j}
	\int dx dx' \, W_j(x) W_j^2(x')
\notag\\
	&\quad\times
	\bigl(
		d_j^\dagger \psi(x)
		+
		d_j \psi^\dagger(x)
	\bigr)
	e^{-i H t_j}
	=
	w_j^2 \hat{W}_j,
\label{sup:W3_2}
\end{align}
with
\begin{equation}
	w_j^2 = \int dx \, W_j^2(x).
\end{equation}
Without the $x$ integrations this
expression would have been left with the complication of
diverging anticommutators that would have required an unnecessary cure, for instance,
through point splitting. From the two results above it follows immediately that
$\hat{W}_j^{2n+1} = w_j^{2n} \hat{W}_j$ and $\hat{W}_j^{2n} = w_j^{2n-2} \hat{W}_j^2$
for integer $n$. Consequently we have
\begin{equation} \label{sup:U_pulse}
	e^{-i \hat{W}_j}
	= \openone - i \alpha_j \hat{W}_j - \beta_j \hat{W}_j^2,
\end{equation}
with
\begin{align}
	\alpha_j &= \sin(w_j)/w_j,
\\
	\beta_j  &= [1-\cos(w_j)]/w_j^2.
\end{align}
This means that the exact form of $e^{-i \hat{W}_j}$ looks like its second order expansion
with renormalised amplitudes. Notably, if we let $w_j \to 0$ then $\alpha_j \to 1$ and
$\beta_j \to 1/2$, matching the second order amplitudes.
Unitarity of Eq.\ \eqref{sup:U_pulse} imposes furthermore that
\begin{equation}\label{sup:trig}
	\alpha_j^2 \hat{W}_j^2 = 2 \beta_j \hat{W}_j^2 - \beta_j^2 \hat{W}_j^4.
\end{equation}
By the same methods that brought Eq.\ \eqref{sup:W3_1} to Eq.\ \eqref{sup:W3_2} we verify that
indeed $\hat{W}_j^4 = w_j^2 \hat{W}_j^2$,
and unitary follows from the trigonometric identity $2 \beta_j - w_j^2 \beta_j^2 = \alpha_j^2$.


\section{Structure of density matrix and correlators}
\label{sec:structure_rho}

Let $\ket{n_1,n_2}$ denote the occupation state of the $d_1$ and $d_2$ levels.
We assume that at time $t=0$ the localised states are in the $\ket{1,0}$ configuration and the fermionic
conductor is in equilibrium. The initial density matrix is thus $\rho(0) = \ket{1,0}\bra{1,0} \otimes \rho_c$,
with $\rho_c$ the conductor's equilibrium density matrix. We apply the first $W_1$ pulse at time $t_1>0$ and the
second $W_2$ pulse at time $t_2>t_1$.
At any time $t>t_2$ the reduced density matrix
takes the form
\begin{equation}
	\rho_d(t)
	= \begin{pmatrix}
		D_0 & 0  & 0   & 0  \\
		0   & A  & C^* & 0  \\
		0   & C  & B   & 0  \\
		0   & 0  & 0   & D_1
	\end{pmatrix},
\end{equation}
spanned in the basis $\{ \ket{0,0}, \ket{1,0}, \ket{0,1}, \ket{1,1} \}$.
The zeros arise from amplitudes that do not preserve the particle number in
the fermionic conductor. (Note that here particle conservation must be considered only along the full Keldysh contour; in real time the number of particles in the conductor is allowed to vary.)
The nonzero amplitudes at times $t > t_2 > t_1$ are given by
\begin{align}
	A &= \bra{1,0} \Tr_c\bigl\{
		e^{-i H t}
		\bigl(\openone - \beta_2 \hat{W}_2^2 \bigr)
		\bigl(\openone - \beta_1 \hat{W}_1^2 \bigr)
		\rho(0)
\notag\\
	&\quad\ \times
		\bigl(\openone - \beta_1 \hat{W}_1^2 \bigr)
		\bigl(\openone - \beta_2 \hat{W}_2^2 \bigr)
		e^{i H t}
	\bigr\} \ket{1,0},
\\
	B &= \alpha_1^2 \alpha_2^2 \,
	     \bra{0,1} \Tr_c\bigl\{
	     e^{-i H t} \hat{W}_2 \hat{W}_1 \rho(0) \hat{W}_1 \hat{W}_2 e^{i H t} \bigr\} \ket{0,1},
\\
	C &= - \alpha_1 \alpha_2 \,
		\bra{1,0} \Tr_c\bigl\{
			e^{-i H t}
			\bigl(\openone - \beta_2 \hat{W}_2^2 \bigr)
			\bigl(\openone - \beta_1 \hat{W}_1^2 \bigr)
\notag\\
	&\quad\ \times
			\rho(0)
			\hat{W}_1 \hat{W}_2
			e^{i H t}
		\bigr\} \ket{0,1},
\\
	D_0 &= \alpha_1^2 \,
		\bra{0,0} \Tr_c\bigl\{
			e^{-i H t}
			\bigl( \openone - \beta_2 \hat{W}_2^2\bigr)
			\hat{W}_1
			\rho(0)
			\hat{W}_1
\notag\\
	&\quad\ \times
			\bigl( \openone - \beta_2 \hat{W}_2^2 \bigr)
			e^{i H t}
		\bigr\} \ket{0,0},
\\
	D_1 &= \alpha_2^2 \,
		\bra{1,1} \Tr_c\bigl\{
			e^{-i H t} \hat{W}_2 \bigl( \openone - \beta_1 \hat{W}_1^2\bigr) \rho(0)
\notag\\
	&\quad\ \times
			\bigl(\openone - \beta_1 \hat{W}_1^2\bigr) \hat{W}_2 e^{i H t}
		\bigr\} \ket{1,1}.
\end{align}
The evaluation of these amplitudes is done by keeping track of which tunnelling transitions are
nonzero on the $d_j$ levels, which gives rise to corresponding $\psi(x_j)$ or $\psi^\dagger(x_j)$
operators. The latter are then rearranged, using the cyclicity of $\Tr_c$ such that standard
correlators $\mean{\dots} = \Tr_c\{ \dots \rho_c\}$ with $\rho_c$, at the far right, are obtained.
Expressions such as $W_j^2 (\psi^\dagger(x_j) \psi(x_j))^2$ are replaced by $w_j^2 \psi^\dagger(x_j) \psi(x_j)$
as shown in Appendix \ref{sec:pulses}. Further simplifications are obtained through identity \eqref{sup:trig}
and the trigonometric relation between $\alpha_j$ and $\beta_j$.

In the expressions below we let $\psi_j = \psi(x_j)$ and use the notations
$h_0 = H|_{V_1=V_2=0}, h_1 = H|_{V_1 \neq 0, V_2 = 0}$, and
$h_{12} = H|_{V_1 \neq 0, V_2 \neq 0}$.
We then obtain
\begin{align}
	A   &= 1 - A_1 - A_2 + A_3 + A_4,\\
	A_1 &= \alpha_1^2 W_1^2 \bmean{e^{i h_0 t_1} \psi_1 \psi_1^\dagger e^{-i h_0 t_1}},\\
	A_2 &= \alpha_2^2 W_2^2 \bmean{e^{i h_0 t_2} \psi_2^\dagger \psi_2 e^{-i h_0 t_2}},\\
	A_3 &= 4 \beta_1 \beta_2 W_1^2 W_2^2
	\notag\\
	    &\quad \times
		  \text{Re}\bmean{ e^{i h_0 t_2} \psi_2^\dagger \psi_2 e^{-i h_0 (t_2-t_1)} \psi_1 \psi_1^\dagger e^{-i h_0 t_1} },\\
	A_4 &= \beta_1^2 \beta_2^2 W_1^4 w_2^2 W_2^2
	\bigl\langle
		e^{i h_0 t_1} \psi_1 \psi_1^\dagger e^{-i h_0 (t_1-t_2)} \psi_2^\dagger \psi_2
	\notag\\
	&\quad \times
		e^{-i h_0 (t_2-t_1)} \psi_1 \psi_1^\dagger e^{-i h_0 t_1}
	\bigr\rangle,
\end{align}
which are all expressions independent of the scattering potentials $V_j$.
The further diagonal entries are
\begin{align}
	B &= \alpha_1^2 \alpha_2^2 W_1^2 W_2^2
	\bigl\langle
		e^{i h_0 t_1} \psi_1 e^{-i h_1 (t_1-t_2)} \psi_2^\dagger\psi_2
	\notag\\
	&\quad\times
	e^{-i h_1 (t_2-t_1)} \psi_1^\dagger e^{-i h_0 t_1}
	\bigr\rangle,
\\
	D_0 &= A_1 - B,\\
	D_1 &= A_2 - A_3 - A_4.
\end{align}
While $D_1$ remains independent of the $V_j$ there is an explicit $V_1$ dependence in $B$ and $D_0$.
However, all time dependence so far is pinned to the pulse times $t_1$ and $t_2$.
The dependence on $t$ (for $t>t_2>t_1$) enters only through the off-diagonal component
\begin{align}
	C &= -\alpha_1 \alpha_2 W_1 W_2
	\bigl\langle
		e^{i h_0 t_1} \psi_1 e^{-i h_1 (t_1-t_2)} \psi_2^\dagger
	\notag\\
	&\quad\times
		e^{-i h_{12}(t_2-t)} e^{-i h_0 t}
	\notag\\
	&\quad\times
		\bigl(
			1
			- \beta_1 W_1^2 e^{i h_0 t_1} \psi_1 \psi_1^\dagger e^{-i h_0 t_1}
		\bigr)
	\notag\\
	&\quad\times
		\bigl(
			1
			- \beta_2 W_2^2 e^{i h_0 t_2} \psi_2^\dagger \psi_2 e^{-i h_0 t_2}
		\bigr)
	\bigr\rangle,
\end{align}
which is also the only term depending on $V_2$ as the latter potential is switched on only
for times $t>t_2$.

The amplitudes $A_{1,2}$ can be evaluated immediately.
Since $h_0$ does not perturb the ground state the correlator
$\mean{ e^{i h_0 t_1} \psi_1 \psi_1^\dagger e^{-i h_0 t_1} }$ equals the local hole density $n_h$.
Likewise $\mean{ e^{i h_0 t_2} \psi_2^\dagger \psi_2 e^{-i h_0 t_2} }$ gives the local particle density $n_p$.
This leads to $A_1 = \alpha_1^2 W_1^2 n_h$ and $A_2 = \alpha_2^2 W_2^2 n_p$.
The remaining amplitudes contain time propagating components and are evaluated through the bosonisation technique.


\section{Bosonisation}
\label{sec:bosonisation}

In the following we focus on one-dimensional (1D) systems such that
the travelling signal remains directed and does not weaken its amplitude by expanding in a higher dimensional space.
This has the additional advantage that we can use the bosonisation technique \cite{Gogolin1998,Giamarchi2007,Shankar2017} which a reliable method
for the explicit evaluation of correlators. Bosonisation allows us furthermore to quantitatively include the renormalisation
of system properties by interactions.
We should emphasise, however, that we choose a 1D system and bosonisation for convenience to provide explicit
analytical results but we do not wish to lose universality by the restriction to the particular pure 1D physics.
Indeed in many cases the interactions in 1D can cause a collective strong coupling response that qualitatively
changes the system's properties. Such physics has been a central theme for 1D systems since many years but it is specific
for this dimensionality. Notable is in particular that backscattering on the impurity causes an interaction-independent
universal long time decay of the standard FES \cite{Gogolin1993}.
Yet here we explicitly exclude such strong coupling physics.
We thus shall use bosonisation in the same spirit as Schotte and Schotte \cite{Schotte1969}
who mapped the radial expansion of a pure s-wave scattering in higher dimensions onto a 1D description solved by bosonisation
and thus could capture in such an elegant way the main many-body features of FES.
For the present 1D description we shall keep nonetheless the fact that modes can travel to the right or the left and use a pure 1D
description, but we either need to assume that backscattering on the impurity does not become relevant for the
described physics (conditions are provided below), or that we deal with a quasi-1D system with sufficient degrees
of freedom in the transverse directions such that the backscattering effect is reduced.
Of course, purely 1D systems without backscattering can be realised experimentally as well such as through chiral quantum Hall edge states
or helical edge states in topological insulators, and for such systems the description below can be applied with only
straightforward adjustments.

The basic condition for bosonisation is that the fermionic band is sufficiently filled such that one can
consider the portions near the Fermi points $\pm k_F$ as two independent bands of right movers $R$ (near $k_F$) and
left movers $L$ (near $-k_F$). The original fermion field operator is then written as
$\psi(x) = e^{i k_F x} \psi_R(x) + e^{-i k_F x} \psi_L(x)$, where $\psi_\nu$ denotes the fermion operator on
the $\nu=R,L$ movers branch. Furthermore the dispersion relation is linearised such
that $\epsilon_{k,\nu} \approx v_F (k-\nu k_F)$,
with the signs $\nu=R=+$ and $\nu=L=-$ replacing the letters $R, L$ where necessary.
The resulting model is known as the Tomonaga-Luttinger model and is described by the Hamiltonian
\begin{align}
	H_{TL} = \sum_\nu \int dx \, \psi_\nu^\dagger (-i \nu v_F) \partial_x \psi_\nu + H_\text{int},
\label{sup:H_TL}
\end{align}
in which we have chosen the chemical potential to be zero such that $H_{TL}$ measures the excitations about the
ground state. The Hamiltonian
$H_\text{int}$ contains the fermion-fermion interactions and can be expressed as
\begin{equation}  \label{sup:H_int}
	H_\text{int} = \sum_{\nu,\nu'} \int dx dx' \, \mathcal{V}(x-x') \psi_\nu^\dagger(x) \psi_{\nu'}^\dagger(x') \psi_{\nu'}(x') \psi_\nu(x),
\end{equation}
with $\mathcal{V}$ the interaction potential.
In $H_\text{int}$ we have omitted terms that couple the $R$ and $L$ movers beyond the written
density-density interaction. Indeed such terms are irrelevant in the renormalisation group sense unless the
fermion density is commensurate with the underlying lattice. We exclude such specific cases here, also in the
spirit of the comments on the choice of a 1D model given above.

The mapping on bosonic degrees of freedom is then a standard procedure (see e.g.\ Refs.\ \cite{Gogolin1998,Giamarchi2007,Shankar2017}
for an in depth discussion),
with the boson fields representing density fluctuations of the $R$ and $L$ movers. The Hamiltonian \eqref{sup:H_TL}
becomes quadratic in the boson fields, and for a sufficiently short ranged (screened) interaction $\mathcal{V}$ such
that the interaction is most pronounced within a range $<\pi/k_F$ all
interactions can be treated as local. The Hamiltonian $H_{TL}$ then becomes a quadratic form described by a $2 \times 2$
matrix for the bosonic $R$ and $L$ fields with the off diagonal terms arising from the $R$ and $L$ density coupling
in $H_\text{int}$. Such a matrix can be immediately diagonalised and the resulting eigenmodes, $\tilde{\phi}_{R,L}$,
describe wave packets that still move only to the right or to the left, although when $\mathcal{V} \neq 0$ both
contain contributions from both original $R$ and $L$ moving density waves. The Hamiltonian is written
accordingly as $H_{TL} = \tilde{H}_R + \tilde{H}_L$ with
\begin{equation} \label{sup:H_nu}
	\tilde{H}_\nu = \int dx \, \frac{v}{4\pi K} \, \bigl(\partial_x \phi_\nu(x)\bigr)^2,
\end{equation}
for $\nu = R,L$. Here $K$ encodes the interaction strength of $\mathcal{V}$, normalised such that $K=1$
corresponds to the non-interacting limit, $0<K<1$ to repulsive interactions and $K>1$ to attractive interactions,
and $v$ is a renormalised Fermi velocity, often equal to $v=v_F/K$.
The eigenmodes $\tilde{\phi}_\nu$ obey the commutation relations
\begin{equation} \label{sup:comm}
	[\partial_x\tilde{\phi}_\nu(x'),\tilde{\phi}_{\nu'}(x)] = 2 i \pi K\nu \delta_{\nu,\nu'} \delta(x-x'),
\end{equation}
such that $\tilde{\phi}_\nu$ and $\partial_x \tilde{\phi}_\nu$ are conjugate boson fields up to a normalisation.
In terms of the eigenmodes the original fermion operators are expressed as
\begin{equation}  \label{sup:psi}
	\psi_\nu(x) = \frac{\eta_\nu}{\sqrt{2\pi a}} e^{-\frac{i}{2}(\nu-K^{-1})\tilde{\phi}_L(x) -\frac{i}{2} (\nu+K^{-1})\tilde{\phi}_R(x)},
\end{equation}
with $a$ a short distance cutoff, typically on the order of the lattice spacing.
The $\eta_\nu$ are Klein factors, operators that lower the overall fermion number by one and
guarantee fermionic exchange statistics. But for the further analysis they produce only unit expectation values and
will be dropped.

The scattering potentials $V_j$ are in their fermionic form given by the Hamiltonian
\begin{equation}
	H_{V} = V_1 \psi^\dagger(x_1) \psi(x_1) d_1 d_1^\dagger + V_2 \psi(x_2) \psi^\dagger(x_2) d_2^\dagger d_2.
\end{equation}
With the splitting into $R$ and $L$ movers $H_V$ has a forward scattering contribution remaining either in the $R$ or
in the $L$ band, and a backscattering contribution transferring between $R$ and $L$. We shall neglect the latter, although
this may seem counter-intuitive as backscattering produces a relevant FES response with a universal time decay that is
independent of $V_j$ \cite{Gogolin1993}. Yet as mentioned above our main aim is to provide a description of the travelling FES
signal and use bosonisation as a convenient tool, but not to be limited to the particularities of the pure 1D response.
In addition, even for the pure 1D case
we should stress that the universal decay is a strong coupling response. It does not set in immediately
but takes a characteristic time $\tau \sim (V_j^b)^{-1} (\xi/V_j^b)^{K/(1-K)} $ to build up before crossing over to the
universal behaviour \cite{Gogolin1993}. Here $V_j^b = |V_j(2k_F)|$ is the backscattering Fourier amplitude of $V_j$.
The time $\tau$ is of significance mostly for strongly interacting systems with $K < 0.7$ at which it can become
very short. However, for typical scales as found in nanowires and not too strong interactions with $K > 0.7$
the value of $\tau$ becomes on the order of microseconds or much larger such that the strong coupling limit
from backscattering is not reached in the time scales governing the described physics otherwise.
The physics then remains perturbative in the backscattering amplitude and has a direct $V_j$ dependence
similar to the effect of forward scattering \cite{Gogolin1993}.
To capture the general effect of FES it is even in this pure 1D situation therefore sufficient
to include only forward scattering.

In this case $H_V$ is expressed in terms of the boson fields as
\begin{align}
	&H_{V}
	= \sum_{\nu} \bigl[
		V_1 \psi_\nu^\dagger(x_1) \psi_\nu(x_1) d_1 d_1^\dagger + V_2 \psi_\nu(x_2) \psi_\nu^\dagger(x_2) d_2^\dagger d_2
	\bigr]
\notag\\
	&=
	\Bigl\{
		\Delta E_1
		-
		\frac{V_1}{2\pi K} \bigl[ \partial_x\tilde{\phi}_L(x_1)+\partial_x\tilde{\phi}_R(x_1) \bigr]
	\Bigr\}
	d_1 d_1^\dagger
\notag\\
	&
	+
	\Bigl\{
		\Delta E_2
		-
		\frac{V_2}{2\pi K} \bigl[ \partial_x\tilde{\phi}_L(x_2)+\partial_x\tilde{\phi}_R(x_2) \bigr]
	\Bigr\}
	d_2^\dagger d_2,
	\label{sup:H_V_bos}
\end{align}
which incorporate the fluctuating parts of the forward scattering.
Here $\Delta E_j = N V_j^2/4\pi v K$, with $N$ the system's particle number, are $d_j$ dependent energy
shifts providing a ground state energy renormalisation by the $V_j$ potentials.
Since Eq.\ \eqref{sup:H_V_bos} is linear in $\tilde{\phi}_\nu$ the total Hamiltonian $H_{TL} + H_V$ can be brought
to the form of $H_{TL}$ by completing the square through a shift in the boson fields,
$\partial_x \tilde{\phi}_\nu - \Delta$, such that the term proportional to
$(\partial_x \tilde{\phi}_\nu) \Delta$ matches $H_V$.
This can be performed on the operator level \cite{Schotte1969} by defining the shift operators
\begin{align}
	\hat{T}_1 &= \exp\left(i \frac{\delta_1}{\pi K} [\tilde{\phi}_R(x_1) -\tilde{\phi}_L(x_1)] d_1 d_1^\dagger\right),\\
	\hat{T}_2 &= \exp\left(i \frac{\delta_2}{\pi K} [\tilde{\phi}_R(x_2) -\tilde{\phi}_L(x_2)] d_2^\dagger d_2\right),
\end{align}
where $\delta_j = 2 K V_j/v$ is the scattering phase shift which for the linearised spectrum matches the
Born approximation \cite{Schotte1969}. Through the commutation relations \eqref{sup:comm} we see that
$H_{TL} + H_V = T_2^\dagger T_1^\dagger H_{TL} T_1 T_2 + \Delta E_1 d_1 d_1^\dagger + \Delta E_2 d_2^\dagger d_2$.
If we let $T_1 = \hat{T}_1|_{d_1 d_1^\dagger =1}$ and $T_2 = \hat{T}_2|_{d_2^\dagger d_2=1}$ then
it follows that
$e^{-i h_1 t} = T_1^\dagger e^{-i h_0 t} T_1 e^{-i \Delta E_1 t}$
and
$e^{-i h_{12} t} = T_2^\dagger T_1^\dagger e^{-i h_0 t} T_1 T_2 e^{-i (\Delta E_1+\Delta E_2) t}$,
which allows us to write the correlators in $\rho_d$ entirely in terms of a time evolution under $h_0$.
For instance, we have
\begin{align}
	C &= -\alpha_1 \alpha_2 W_1 W_2
	\bigl\langle
		e^{i h_0 t_1} \psi_1 T_1^\dagger e^{-i (h_0+\Delta E_1) (t_1-t_2)} T_1 \psi_2^\dagger
	\notag\\
	&\quad\times
		T_2^\dagger T_1^\dagger e^{-i (h_0+\Delta E_1+\Delta E_2)(t_2-t)} T_1 T_2 e^{-i h_0 t}
	\notag\\
	&\quad\times
		\bigl(
			1
			- \beta_1 W_1^2 e^{i h_0 t_1} \psi_1 \psi_1^\dagger e^{-i h_0 t_1}
		\bigr)
	\notag\\
	&\quad\times
		\bigl(
			1
			- \beta_2 W_2^2 e^{i h_0 t_2} \psi_2^\dagger \psi_2 e^{-i h_0 t_2}
		\bigr)
	\bigr\rangle.
\end{align}
Inserting the time dependence $O(t) = e^{i h_0 t} O e^{-i h_0 t}$ for any operator $O$
the latter expression can be rewritten as
\begin{align}
	C &= -\alpha_1 \alpha_2 W_1 W_2
	e^{-i \Delta E_1 (t_1-t)}
	e^{-i \Delta E_2(t_2-t)}
	\notag\\
	&\quad\times
 	\bigl\langle
		\psi_1(t_1) T_1^\dagger(t_1) T_1(t_2) \psi_2^\dagger(t_2)
		T_2^\dagger(t_2) T_1^\dagger(t_2) T_1(t) T_2(t)
	\notag\\
	&\quad\times
		\bigl(
			1
			- \beta_1 W_1^2 \psi_1(t_1) \psi_1^\dagger(t_1)
		\bigr)
	\notag\\
	&\quad\times
		\bigl(
			1
			- \beta_2 W_2^2 \psi_2^\dagger(t_2) \psi_2(t_2)
		\bigr)
	\bigr\rangle.
\end{align}
Similarly we find
\begin{align}
	B &= \alpha_1^2 \alpha_2^2 W_1^2 W_2^2
	\bigl\langle
		\psi_1(t_1) T_1^\dagger(t_1) T_1(t_2) \psi_2^\dagger(t_2) \psi_2(t_2)
	\notag\\
	&\quad\times
	T_1^\dagger(t_2) T_1(t_1) \psi_1^\dagger(t_1)
	\bigr\rangle,
\end{align}
and corresponding expressions without $T_j$ operators for all other amplitudes in $\rho_d$.
In all these expressions the fermion operators are replaced by Eq.\ \eqref{sup:psi} and we note that
\cite{Gogolin1998,Giamarchi2007}
\begin{align}
	\psi^\dagger(x) \psi(x)
	&= n_p - \frac{1}{2\pi} [\partial_x \tilde{\phi}_L(x) + \partial_x\tilde{\phi}_R(x)]
\notag\\
	&+ \frac{1}{2\pi a}
	\left( e^{i [ \tilde{\phi}_L(x) + \tilde{\phi}_R(x)]} + \text{h.c.}\right),
\end{align}
and $\psi(x)\psi^\dagger(x) = (n_p+n_h)-\psi^\dagger(x) \psi(x)$, where the densities $n_p$ and $n_h$
regularise the divergences from the delta function of the anticommutator.

The final evaluation of all correlators is done by using the identity
$\langle \prod_i \exp(\lambda_i \tilde{\phi}_i) \rangle =
\exp(\sum_{i<j} \lambda_i\lambda_j  G_{ij})$
valid for any theory with a quadratic bosonic Hamiltonian where $\tilde{\phi}_i = \tilde{\phi}_{\nu_i}(x_i,t_i)$
and \cite{Shankar2017}
\begin{align}
	&G_{ij}
	=
	\langle \tilde{\phi}_i \tilde{\phi}_j - (\tilde{\phi}_i^2 + \tilde{\phi}_j^2 )/2 \rangle
\notag\\
	&= - K \delta_{\nu_i,\nu_j} \ln\left[\frac{a - i\nu_i(x_i-x_j) + i v (t_i-t_j)}{a}\right].
\end{align}
In the latter equation we use the zero temperature $T=0$ limit which is applicable as long as all
considered time scales are shorter than the thermal time $\tau_T = 2\pi/k_B T$.
For density-density correlators involving products of gradients we have
\begin{align}
	\langle \partial_x \tilde{\phi}_i \partial_x \tilde{\phi}_j \rangle
	=
	\frac{K \delta_{\nu_i,\nu_j} }{[a - i\nu_i(x_i-x_j) + i v (t_i-t_j)]^2}.
\label{eq:grad-grad}
\end{align}
A subtlety arises from the term $\psi_1(t_1) \psi_2^\dagger(t_2) \psi_2(t_2) \psi_1^\dagger(t_1)$
in $B$ which must remain real and the standard point splitting method of bosonisation is ambiguous.
But there is no ambiguity in the noninteracting $K=1, \delta_j=0$ limit in which the evaluation of the
fermionic amplitude is a basic application of Wick's theorem.
From continuity with this result we find that in this case the correlators of the form of
Eq.\ \eqref{eq:grad-grad} must instead be given by
$-K \delta_{\nu_i,\nu_j}/\{a^2 + [\nu_i(x_i-x_j) - v (t_i-t_j)]^2\}$.

The final results for the amplitudes $A, B, C, D_{0,1}$ are then obtained straightforwardly
but require a good bookkeeping as they consist of products of many factors of the form
$\left\{[a - i\nu_i(x_i-x_j) + i v (t_i-t_j)]/a\right\}^{\gamma_{ij}}$ with the various
exponents $\gamma_{ij}$ arising from the $\psi$ and $T$ operators. We shall write out explicitly only
the leading part of the amplitude $C$, without the higher order contributions proportional to
$\beta_{1,2}$, as the discussion in the main text focuses on the latter.
We have
\begin{align}
	C
	&=
	-\alpha_1 \alpha_2 W_1 W_2
	e^{-i \Delta E_1 (t_1-t)}
	e^{-i \Delta E_2(t_2-t)}
\notag\\
	&\times
	\bigl( e^{-i k_F(x_1-x_2)} C_- + e^{+i k_F(x_1-x_2)} C_+ \bigr) / 2\pi a,
\end{align}
in which the amplitudes $C_\nu$, for $\nu = \pm$ (denoting $R$ and $L$ movers, respectively), are products of the power laws arising from the
multiple contractions between the boson fields.
With $g_{x,t} = (a - i x + i v t)/a$ we have
\begin{align}
	&C_\nu
	=
	g_{0,       t_1-t}^{   \frac{\delta_1}{\pi K}-\frac{2\delta_1^2}{\pi^2 K}}
	\
	g_{0,       t_2-t}^{  -\frac{\delta_2}{\pi K}-\frac{2\delta_2^2}{\pi^2 K}}
\notag\\&\times
	g_{x_1-x_2, 0}^{      -\frac{\nu \delta_1}{\pi} + \frac{\delta_1 \delta_2}{\pi^2 K}}
	\
	g_{x_2-x_1, 0}^{       \frac{\nu \delta_1}{\pi} + \frac{\delta_1 \delta_2}{\pi^2 K}}
\notag\\&\times
	g_{x_1-x_2, t_1-t_2}^{-\frac{(1+K \nu)^2}{4 K}+\frac{(\delta_1-\delta_2)(1+K \nu)}{2\pi K} + \frac{\delta_1 \delta_2}{\pi^2 K}}
	g_{x_1-x_2, t_1-t}^{   \frac{\delta_2(1+K \nu)}{2\pi K} - \frac{\delta_1 \delta_2}{\pi^2 K}}
\notag\\&\times
	g_{x_2-x_1, t_1-t_2}^{-\frac{(1-K \nu)^2}{4 K}+\frac{(\delta_1-\delta_2)(1-K \nu)}{2\pi K} + \frac{\delta_1 \delta_2}{\pi^2 K}}
	g_{x_2-x_1, t_1-t}^{   \frac{\delta_2(1-K \nu)}{2\pi K} - \frac{\delta_1 \delta_2}{\pi^2 K}}
\notag\\&\times
	g_{x_1-x_2, t_2-t}^{  -\frac{\delta_1(1-K \nu)}{2\pi K} - \frac{\delta_1 \delta_2}{\pi^2 K}}
	g_{x_2-x_1, t_2-t}^{-\frac{\delta_1(1+K \nu)}{2\pi K} - \frac{\delta_1 \delta_2}{\pi^2 K}}.
\label{eq:C_nu_full}
\end{align}
The first two $x_j$ independent factors arise only from the OC. The next two time independent factors compensate in
amplitude but provide a phase to the coefficient. The further terms encode the full spatio-temporal spread
of the FES signal, with peaks whenever in $g_{x,t}$ the condition $x-vt=0$ is met.
At $t=t_2$ this expression simplifies to
\begin{align}
	&C_\nu
	=
	g_{0,       t_1-t_2}^{   \frac{\delta_1}{\pi K}-\frac{2\delta_1^2}{\pi^2 K}}
	g_{x_1-x_2, 0}^{  -\frac{\delta_1(1+K \nu)}{2\pi K}}
	g_{x_2-x_1, 0}^{-\frac{\delta_1(1-K \nu)}{2\pi K}}
\notag\\&\times
	g_{x_1-x_2, t_1-t_2}^{-\frac{(1+K \nu)^2}{4 K}+\frac{\delta_1(1+K \nu)}{2\pi K}}
	g_{x_2-x_1, t_1-t_2}^{-\frac{(1-K \nu)^2}{4 K}+\frac{\delta_1(1-K \nu)}{2\pi K}},
\label{eq:C_nu_t2}
\end{align}
in which all dependence on $\delta_2$ drops out.

%

\end{document}